\documentclass[12pt]{article}
\def\Journal#1#2#3#4#5#6{(#5)  {#1} {\bf #2} #3}

\def\CQG{\em Class. Quantum Grav.}

\def\GRG{\em Gen. Rel. Grav.}

\def\PR{\em Phys. Rev.}

\def\JMP{\em J. Math. Phys.}

\def\CMP{\em Commun. Math. Phys.}

\def\AIHP{\em Ann. Inst. H. Poincar\'e}

\def\be{\begin{equation}}
\def\ee{\end{equation}}
\def\bea{\begin{eqnarray}}
\def\eea{\end{eqnarray}}
\def\bean{\begin{eqnarray*}}
\def\eean{\end{eqnarray*}}

\newcommand{\bm}[1]{\mbox{\boldmath $#1$}}

\def\stat{\vec s}
\def\fstat{\bm{s}}
\def\st{s}
\def\pp{\varphi}

\def\Parz{\frac{\partial}{\partial z}}

\def\Part{\frac{\partial}{\partial t}}
\def\Parp{\frac{\partial}{\partial \varphi}}

\def\R{{\rm I\!R}}
\def\d{{\rm \mbox{d}}}
\def\g{{\rm \mbox{g}}}

\def\espaitemps{({\cal V},\g)}
\def\varietat{{\cal V}}

\def\fin{\hfill \rule{2.5mm}{2.5mm}\\ \vspace{0mm}}
\def\finn{\hfill \rule{2.5mm}{2.5mm}}

\def\proof{\noindent{\em Proof.\/}\hspace{3mm}}

\newtheorem{teorem}{Theorem}

\newtheorem{propos}{Proposition}

\newtheorem{definicio}{Definition}
\newtheorem{corollary}{Corollary}[propos]

\newcounter{eqlletra}

\def\subsubsection#1{\vspace{8mm} \noindent {\bf #1}  \vspace{3mm}}
\def\subsubsubsection#1{\vspace{8mm} \noindent {\it #1}  \vspace{3mm}}

\begin{document}
\title{On the definition of cylindrical symmetry}
\author{Jaume Carot$^\natural$,
Jos\'e M. M. Senovilla\thanks{Also at
Laboratori de F\'{\i}sica Matem\`atica,
Societat Catalana de F\'{\i}sica, IEC, Barcelona.}$^\ast$$^\flat$ and
Ra\"ul Vera\footnotemark[1]$^\ast$\\
$\natural$ Departament de F\'{\i}sica, Universitat de les Illes Balears,\\
E-07071 Palma de Mallorca, Spain.\\
$\ast$ Departament de F\'{\i}sica Fonamental, Universitat de Barcelona\\
 Diagonal 647, 08028 Barcelona, Catalonia, Spain.\\
$\flat$ Fisika Teorikoaren Saila, Euskal Herriko Unibertsitatea,\\
644 P.K., 48080 Bilbo, Spain.}
\maketitle
\begin{abstract}
The standard definition of cylindrical symmetry in General Relativity
is reviewed. Taking the view that axial symmetry is an essential pre-requisite
for cylindrical symmetry, it is argued that the requirement of
orthogonal transitivity of the isometry group should be dropped, this
leading to a new, more general definition of cylindrical symmetry.
Stationarity and staticity in cylindrically symmetric spacetimes are then
defined, and these issues are analysed in connection with
orthogonal transitivity, thus proving some new results on the
structure of the isometry group for this class of spacetimes.
\end{abstract}

\section{Introduction}

The purpose of this paper is to discuss the standard definition
of cylindrically symmetric spacetimes and give
some remarks on its possible generalizations.
In particular, the assumptions which are usually
made but are not necessary are pointed out, and the
results herein presented will also be valid in some
more general situations. Special attention is devoted
to the
{\it stationary and static\/}
cylindrically symmetric cases.

The intuitive idea about cylindrical symmetry is very clear.
However, there are some subtleties which deserve attention in general
relativity. Just as an example we can remember that there are cases
in which the axis of symmetry is spatially closed
(a closed RW geometry, for instance), which may not seem in accordance with
the standard view of a ``cylinder''.
Our main assumption is that
there is an axial Killing vector and that at least
part of its axis of symmetry belongs to the spacetime.
 This will be absolutely essential for all our results.
Of course, we could also consider situations where the axis
of symmetry is completely absent, as
for instance, when treating the exterior field for a cylindrical
source. The axis is inside the source and thus the exterior
field could be just any spacetime with a Killing vector having
closed orbits. These Killing vectors can be obtained by identifying
points in spacetimes with a spacelike symmetry, see also
\cite{MCLLUMcil}.
Nevertheless, our assumption is
justified because any globally defined cylindrically symmetric
spacetime will usually contain the axis.

Keeping the above assumption in mind, we need another spacelike symmetry
such that the orbits of the $G_2$ group are {\em locally\/}
cylinders, which must be assumed
to be spacelike.
The existence of 2-surfaces orthogonal to the
group orbits is an extra assumption, not necessary
for the definition of cylindrical symmetry, as we will see in a
well known example, although in certain
situations it holds as a consequence of
the form of the Ricci tensor and the existence of the axis of symmetry.
In summary, the basic ingredient for the
cylindrical symmetry is a $G_2$ on $S_2$ group of motions containing
an axial symmetry with the axis present in the given spacetime.

\section{Axial and cylindrical symmetry}

The purpose of this section is to review the definition of axial
symmetry along with its  associated basic geometrical features,
and to put forward and discuss a definition of cylindrical
symmetry, exploring its consequences.

Regarding axial symmetry, one has the following definition
(see \cite{maseaxconf,Marctesi}):

\begin{definicio}
A spacetime $\espaitemps$ is said to have axial symmetry if and only if
there is an effective realization of the one dimensional torus $T$
into $\varietat$ that is an isometry and such that its set of fixed points
is non-empty.
\label{def:axial}
\end{definicio}

Notice that Definition \ref{def:axial} implicitly assumes that there
exists at least one fixed point
(i.e. points that remain invariant under the action of the group)
in $\espaitemps$.
In fact, it can be proven that the set of fixed points
must be an autoparallel, 2-dimensional timelike surface.
This surface is the axis of symmetry and will henceforth be  denoted as $W_2$
\cite{maseaxconf,Marctesi,carter70,chuscomm}. In previous standard
definitions
the axis was assumed to be a 2-dimensional surface \cite{carter70,chuscomm},
but, as we have just mentioned, this is necessarily so and therefore needs
not be assumed as an extra requirement in the definition af axial
symmetry \cite{maseaxconf,Marctesi}.

Furthermore, it can be shown \cite{maseaxconf,Marctesi} that
the infinitesimal
generator $\vec \xi$ of the axial symmetry is spacelike in a neighbourhood
of the axis, and that the so-called elementary flatness condition
holds \cite{maseaxconf,KRAM}, that is :
\be
\left.\frac{\nabla_\rho (\xi_\alpha \xi^\alpha)
\nabla^\rho (\xi_\beta \xi^\beta)}{4 \xi_\rho \xi^\rho}
\right|_{W_2}
\longrightarrow 1.
\label{eq:condregaxis}
\ee
This condition
ensures the standard 2$\pi$-periodicity of the axial coordinate
near the axis.

Further fundamental results concern the relation of the Killing
vector $\vec \xi$ with other vector fields, and in particular
with different isometry generators.
We refer to \cite{maseaxconf,Marctesi,carter70,chuscomm}
for the proofs:

\begin{teorem}
Let $\vec v$ be a vector field in an axisymmetric spacetime and $q\in W_2$.
\begin{enumerate}
\item
$\vec v|_q$ is tangent to the axis at $q$ iff $[\vec v,\vec \xi]|_q=0$.
\item
$\vec v|_q\;(\neq 0)$ is normal to the axis at $q$ iff $\vec v|_q$
and $[\vec v,\vec \xi]|_q$ are linearly independent vectors and
$[[\vec v,\vec \xi],\vec \xi]|_q$ depends linearly on the previous.
\item
$\vec v$ is neither tangent nor normal to the axis at $q$ iff
$\vec v|_q$, $[\vec v,\vec \xi]|_q$ and $[[\vec v,\vec \xi],\vec \xi]|_q$
are linearly independent vectors and
$[[[\vec v,\vec \xi],\vec \xi],\vec \xi]|_q$ depends linearly on the
previous two.\finn
\end{enumerate}
\label{teo:4.2}
\end{teorem}

\begin{teorem}
In an axially symmetric spacetime, if $\vec \lambda$ is a
Killing vector field tangent to the axis of symmetry for all $q\in W_2$,
then

\begin{center}$[\vec \xi,\vec \lambda]=0.$\end{center}
\vspace{-28pt}\finn
\label{teo:5.2}
\end{teorem}

\begin{propos}
In an axisymmetric spacetime, let $\vec \lambda$ be a Killing
vector field which does not commute with $\vec \xi$. If at some
point $q$ of the axis $\vec \lambda|_q$ is not normal
to $W_2$, then there always exists another
Killing vector field given by
$\vec \lambda+[[\vec \lambda,\vec \xi],\vec \xi]$ that commutes with $\vec
\xi$, and is therefore tangent to the axis.\finn
\label{teo:prop5.3}
\end{propos}

It should be noticed that all the results above apply also to conformal Killing
vector fields \cite{maseaxconf,Marctesi}.

\vspace{5mm}
Let us next consider the definition of cylindrical symmetry.
In addition to the existence of two spacelike Killing vector fields,
$\vec \xi$ and  $\vec \eta$, one of which, say  $\vec \xi$, is taken to
generate an axial symmetry, it has been usually assumed that both Killing
vectors commute and that the $G_2$ acts orthogonally transitively.
With regard to the  assumption of commutativity, from
Proposition \ref{teo:prop5.3} it is clear that
the existence of a Killing vector field
that is not orthogonal to $W_2$ at some point would suffice.
However, not even this assumption is actually necessary due to the
following result:

\begin{propos}
In an axially symmetric spacetime, if there is another
Killing vector $\vec \lambda$ which generates
with $\vec \xi$ a $G_2$ group, then both Killing vectors
commute, thus generating an Abelian $G_2$ group.
\label{teo:prop2}
\end{propos}

\proof
If $\vec \lambda|_q$ is not orthogonal to the axis
for a given point $q\in W_2$, then
from Proposition \ref{teo:prop5.3} we have that the
vector field $\vec \lambda+[[\vec \lambda,\vec \xi],\vec \xi]$,
which belongs to the same $G_2$ group, commutes with $\vec \xi$,
leading to an Abelian $G_2$ group.
Suppose then that $\vec \lambda$ is orthogonal
to $W_2$ at all its points. From Theorem \ref{teo:4.2} point 2
it follows that another independent Killing vector field given by
$\vec \lambda'\equiv [\vec \xi,\vec \lambda]$ exists;
but this leads to a contradiction because we are under the
assumption that $\vec \xi$ and $\vec \lambda$ generate a group
of isometries.\fin


The assumption on the existence of 2-surfaces orthogonal to the
group orbits (see, for instance \cite{MCLLUMcil,KRAM,MEL,T})
is not necessary for the definition of cylindrical symmetry
nor a consequence of it,
as we will see in an explicit example below,
although its justification
would come mainly from three different sorts of reasons.
The first one concerns the invertibility of the
$G_2$ group, which is equivalent to its
orthogonal transitivity \cite{carter69}.
The second corresponds to the considerations
given by Melvin \cite{MEL,T}
about the invariance under reflection in planes
containing the axis and perpendicular to it (this is
explicitly used in the definition of the whole cylindrical
symmetry, that is, such that $\vec \xi$ and $\vec \eta$ are also
mutually orthogonal).
This is, in fact, equivalent to demanding the invertibility
of each of the one-parameter subgroups forming the Abelian
$G_2$, and thus it is a particular case of the first assumption.
The previous reasonings are geometrical in nature, while
the third is based on results
concerning the form of the Ricci tensor for
some interesting material contents, such as
$\Lambda$-terms (including vacuum) and
perfect fluids whose velocity vector $\vec u$
is orthogonal to the group orbits, since in those cases it can be
shown, see \cite{chuscomm,KRAM,carter69}, on account of
the vanishing of $\vec \xi$ at $W_2$, that orthogonal transitivity follows.

The possible definition for cylindrically symmetric spacetimes,
avoiding complementary assumptions, could thus be:

\begin{definicio}
A spacetime $\espaitemps$ is cylindrically symmetric
if and only if it admits a $G_2$ on $S_2$
group of isometries
containing an axial symmetry.
\label{def:cilin}
\end{definicio}
The line-element of cylindrically symmetric spacetimes
corresponds then to that of
the Abelian $G_2$ on $S_2$ spacetimes \cite{wainwright},
since this definition automatically implies that the
$G_2$ group must be Abelian as follows from Proposition \ref{teo:prop2} above.
Orthogonal transitivity is then left as an extra assumption,
taking into account that, as was already mentioned, it
follows directly in some important cases
from the structure of the Ricci tensor and
the existence of an axis.


The above definition is inspired by the intuitive
idea of cylindrically symmetric spacetimes as those
containing spatial cylinders, which are just
$S^1\times V_1$ spacelike surfaces {\em with a flat metric}.
Here, by $V_1$ we mean any of $S^1$ or $\R$ spaces,
that is, we consider not only spatially infinite axis
of symmetry, but also spatially finite axis
which may appear (as in a closed RW geometry).
Notice that these $S^1\times S^1$ surfaces are {\em not}
standard toruses since the first fundamental form of a standard
torus is non-flat.

Non-orthogonally transitive Abelian $G_2$ on $S_2$ spacetimes
with an axial symmetry (metrics of types A(i) and A(ii)
in Wainwright's classification \cite{wainwright})
must be considered
as cylindrically symmetric as they contain
a two-parameter family of imbedded spatial cylinders.
A well-known explicit
example is given by the dust spacetime with line-element
((20.13) in \cite{KRAM})
\be
\d s^2=e^{-a^2\rho^2}\left( \d \rho^2+\d z^2\right)+\rho^2 \d \pp^2
-\left(\d t+a\rho^2 \d \pp\right)^2,
\label{eq:2013}
\ee
which belongs to the van Stockum class of stationary axisymmetric
dust spacetimes \cite{vanStockum},
whose $G_2$ on $S_2$ group is non-orthogonally
transitive. The fluid flow (tangent to $\partial/\partial t$)
is not orthogonal to the group orbits, as otherwise the
orthogonal transitivity would follow necessarily from the
perfect-fluid form of the Ricci tensor and the existence
of the axis of symmetry, see above.
The spacelike character of the axial Killing vector
is ensured in a region around the axis. The spacetime can
be matched then to a static vacuum metric \cite{KRAM}.
The surfaces given by $\{t=\mbox{const.}, \rho=\mbox{const.}\}$
constitute the two-parameter family of imbedded spatial cylinders,
which are rigidly rotating around the axis of symmetry ($\rho=0$).

This situation regarding the orthogonal transitivity
in cylindrical symmetry is clearly in contrast with the case of
spherical symmetry, where the existence of surfaces
orthogonal to the group orbits is {\em geometrically\/} ensured
\cite{KRAM,schmidt67}.


\section{Stationary and static cylindrically symmetric spacetimes}

Once we have discussed cylindrical symmetry,
we can proceed further and study the definitions of
both stationary and static cylindrically symmetric spacetimes.
Stationarity implies the existence of an additional isometry
which is generated by a timelike Killing vector
field (that is integrable in the static case).
The first consequence is that, since a timelike vector field
cannot be orthogonal to $W_2$ anywhere, Proposition \ref{teo:prop5.3}
and Theorem \ref{teo:5.2} imply the existence of a Killing vector
$\vec \zeta$ such that $[\vec \xi,\vec \zeta]=0$ which can be
checked to be timelike in the region where the original one
was timelike \cite{carter70,chuscomm,maseaxconf}.
Therefore, at this stage, we have that
the group structure
of stationary cylindrically symmetric spacetimes
is
a $G_3$ on $T_3$ group of isometries generated by
two spacelike Killing vectors $\vec \xi$ and $\vec \eta$,
and a timelike Killing vector $\vec \zeta$,\footnote{Although
the existence of such a timelike Killing vector field in a spacetime
with a $G_3$ on $T_3$ is not ensured globally, it is certainly true
in some open neighbourhood of any given point.}
such that
$\vec \xi$ commutes with both $\vec \eta$ and $\vec \zeta$,
\be
[\vec \xi,\vec \eta]=0,\hspace{1cm}
[\vec \xi,\vec \zeta]=0.\label{xietazeta}
\label{eq:comm}
\ee

In the static case we must further impose the existence
of an integrable timelike Killing vector $\stat$.
It should be noticed that in the static case,
$\stat$ does not necessarily coincide with $\vec \zeta$ in principle.

Notice that the definition of stationary cylindrically symmetric spacetimes
which appears in \cite{KRAM} includes the extra assumption
$[\vec \eta,\vec \zeta]=0$, apart from the orthogonal
transitivity on the $G_2$ on $S_2$ assumed in the
definition of cylindrical symmetry in this reference.
However, as we will see later in the next
section, the assumption $[\vec \eta,\vec \zeta]=0$
together with orthogonal transitivity of the $G_2$ on $S_2$ subgroup
implies, in fact, staticity.
Therefore, in order to look for actual stationary (non-static)
models of these characterictics, one of the two extra assumptions must
be dropped.

As a matter of fact, in \cite{KRAM} the extra assumption
$[\vec \eta,\vec \zeta]=0$ is maintained instead
of the orthogonal transitivity on the $G_2$ on $S_2$ subgroup, stating
essentially that (the phrasing is ours)
``stationary cylindrically symmetric spacetimes are those
admitting an Abelian $G_3$ on $T_3$ group of isometries
containing
a $G_2$ on $S_2$ subgroup with an axial symmetry''.
Of course, this is not coherent with the assumption of
orthogonal transitivity in the definition of
cylindrical symmetry that appears in the same reference.
Indeed, the metric (\ref{eq:2013}) is presented in \cite{KRAM}
as an example of stationary cylindrically symmetric spacetime,
although its $G_2$ on $S_2$ group does not act orthogonally transitively.
Let us remark that the metrics appearing in Section 20.2
of \cite{KRAM}, which are presented as stationary cylindrically
symmetric {\it vacuum} solutions, also possess a $G_2$ on $S_2$ which
is not orthogonally transitive, but this necessarily implies
that the axial symmetry cannot be well-defined in these {\em vacuum}
spacetimes, as we have mentioned in the previous section.
Nevertheless, all these vacuum examples could be matched to another
cylindrically symmetric spacetime with the axis included,
which would be then considered as the source of the exterior
vacuum spacetime, so that the axis of symmetry would not
be present in the vacuum region.

In the next section we will focus on the assumption that
the $G_2$ on $S_2$ acts orthogonally transitively, which
will give some results concerning the group structures
and the form of the line-elements.
This study has also a clear motivation, since
in some relevant afore mentioned cases (including vacuum)
this assumption is a direct consequence.

\section{Stationarity, staticity and orthogonal transitivity
in cylindrically symmetric spacetimes}

Let us assume now that
$\vec \xi$ and $\vec \eta$ generate
an Abelian subgroup $G_2$ whose orbits $S_2$ admit orthogonal
surfaces, i.e.:
$$
\bm{\xi}\wedge \bm{\eta} \wedge \d \bm{\xi}=0,\hspace{1cm}
\bm{\xi}\wedge \bm{\eta} \wedge \d \bm{\eta}=0.
$$
It is straightforward to show that there are four non-isomorphic
algebraic structures for the $G_3$ group
generated by $\{\vec \xi,\vec \zeta,\vec \eta\}$
satisfying (\ref{eq:comm})
\cite{maseaxconf,Marctesi,cacosi,alitesi,ali,MATHO},
and taking into account that $\vec \xi$ vanishes on the axis,
the remaining commutator can then be expressed,
in each case and without loss of generality, as

\begin{enumerate}
\item{Abelian Case:} (Bianchi I) $[\vec \eta,\vec \zeta]=0$,
\item Case I: (Bianchi III) $[\vec \eta,\vec \zeta]=b\vec \zeta$,
\item Case II: (Bianchi III) $[\vec \eta,\vec \zeta]=c\vec \eta$,
\item Case III: (Bianchi II) $[\vec \eta,\vec \zeta]=d\vec \xi$,
\end{enumerate}
where $b,c,d$ are constants. Some of these constants
could have been set equal to 1 by re-scaling conveniently the Killing vectors,
but we choose not to do so because they can carry physical units.
Notice that the above algebraic structure does not depend
on the timelike or spacelike character of the Killing vetor
field $\vec \zeta$.
Now, taking into account that
$\vec \xi$ and $\vec \eta$ span a subgroup which acts orthogonally
transitively and using the fact that we want
the $G_3$ group acting on $T_3$, so that the projection
of the globally defined Killing vector field $\vec \zeta$
onto the surfaces {\em orthogonal} to the orbits generated by
the $G_2$ subgroup $\{\vec \xi,\vec \eta\}$ is necessarily
timelike, it follows that we can choose coordinates
$\{t,x,\pp,z\}$ such that
\be
\vec \xi=\Parp,\hspace{1cm}
\vec \eta=\Parz,
\label{eq:xieta}
\ee
and the line-elements for each of the above algebras can then be written
as follows (see \cite{Marctesi,cacosi,alitesi,ali,MATHO}):

\subsubsection{Abelian Case:}

The line-element is given by
\be
\d s^2=\frac{1}{S^2(x)}\left[-\d t^2 +\d x^2+\frac{Q^2(x)}{F(x)}\d \pp^2+
F(x)\left(\d z+W(x)\d \pp\right)^2\right],
\label{eq:abel}
\ee
where $S$, $Q$, $F$ and $W$ are arbitrary functions of $x$
and the Killing vector $\vec \zeta$ has the following expression
\[
\vec \zeta=\Part.
\]
In this case we can choose $\vec\zeta=\stat$
because $\vec\zeta$ is already an integrable timelike
Killing vector field and thus we have a static cylindrically
symmetric spacetime.
This indirectly proves the following:
\begin{propos}

Given an Abelian $G_{3}$ on $T_{3}$ containing a subgroup  $G_2$ on
$S_2$ acting orthogonally transitively,
there always exists an integrable timelike Killing
vector field.

\finn
\end{propos}

This applies, in fact, for $G_2$ on $V_2$ and an additional conformal
Killing vector field with the `opposite' character
\cite{Marctesi,MATHO}.
Therefore, we have

\begin{corollary}

A (non-static) stationary
spacetime cannot contain an orthogonally transitive Abelian $G_2$ on $S_2$
subgroup whenever the $G_3$ group containing these symmetries is Abelian.
\end{corollary}

\subsubsection{Case I:}

The line-element takes now the form
\[
\d s^2=\frac{1}{S^2(x)}\left[-\d t^2 +\d x^2+b^2M^2(t) \d z^2+
L^2(x)\left(\d \pp+b N(t) \d z\right)^2\right],
\]
where $M$ and $N$ are functions of $t$ satisfying
$M_{,t}^2=1+\alpha M^2$ with $M_{,t}\neq 0$, $N_{,t}=\omega M$,
$\alpha,\omega$ are constants, and
$L$ is an arbitrary function of $x$.
The Killing vector $\vec \zeta$ reads
\be
\vec \zeta=e^{bz}\left(-\frac{1}{b}\frac{M_{,t}}{M}\Parz
+\left(N\frac{M_{,t}}{M}-\omega M\right)\Parp +\Part\right).
\label{eq:stat3}
\ee
This vector field is timelike if $\alpha+L^2\omega^2<0$.

\subsubsection{Case II:}

In this case the line-element has the following expression
\[
\d s^2=\frac{1}{S^2(x)}\left[-\d t^2 +\d x^2+\frac{Q^2(x)}{F(x)}\d \pp^2+
F(x)\left(e^{-ct}\d z+W(x)\d \pp\right)^2\right],
\]
and we have then
\[
\vec \zeta=c z \Parz+\Part,
\]
which is timelike whenever $c^2z^2Fe^{-2ct}-1<0$.

\subsubsection{Case III:}

The line-element reads now
\[
\d s^2=\frac{1}{S^2(x)}\left[-\d t^2 +\d x^2+F(x)\d z^2+
\frac{Q^2(x)}{F(x)}\left(\d \pp+(W(x)-td)\d z\right)^2\right],
\]
and $\vec \zeta$ is given by
\[
\vec \zeta=zd\Parp+\Part,
\]
being timelike in the region $z^2 d^2Q^2-F<0$.

The only cases in which the timelike character
of the Killing vector $\vec \zeta$ is ensured all over the spacetime
are the Abelian case
and also in  Case I, once the function $L(x)$ has been
chosen appropriately.

In order to see whether or not a globally defined timelike Killing vector
field exists in the non-Abelian cases, we consider a general Killing vector
$\stat$
not contained in the $G_2$, i.e.:
\be
\stat=\vec \zeta + A \vec \xi+ B\vec \eta,
\label{eq:stat}
\ee
where $A$ and $B$ are arbitrary constants,
and compute its modulus in each case. It follows:
\bea
\mbox{Case I:}&&(\stat\cdot \stat)=
\frac{1}{S^2}\left\{e^{2bz}M^2(\alpha+L^2\omega^2)-
2Me^{bz}\left[A\omega L^2+Bb(\omega NL^2+M_{,t})\right]\right.\nonumber\\
&&\hspace{2cm}\left.+L^2\left(A+b B N\right)^2+B^2b^2M^2\right\}
\label{eq:sscasei}\\
\mbox{Case II:}&&(\stat\cdot \stat)=
\frac{1}{S^2}\left\{-1+A^2\frac{Q^2}{F}+F\left(cze^{ct}+AW\right)^2\right\},
\nonumber\\
\mbox{Case III:}&&(\stat\cdot \stat)=
\frac{1}{S^2}\left\{-1+B^2F+\frac{Q^2}{F}\left(zd+B(W-td)\right)^2\right\}.
\nonumber
\eea
From the above expressions it is immediate to see that in Cases II and
III, and for any given functions of $x$ and constants $A$ and $B$,
we can reach points where $(\stat\cdot \stat)>0$
whenever the coordinate $z$ can reach any value in $(-\infty,\infty)$.
Therefore, stationary spacetimes with a globally defined
timelike Killing vector field whose axis of symmetry
extend indefinitely in the $z$--coordinate can only exist in the Abelian
case or in some situations of the Case I.

Let us next investigate the  existence of integrable
Killing vectors in Cases I, II and III. If one such  vector field
outside the $G_2$ group exists, $\stat$, it must be of the form  given by
(\ref{eq:stat}) although it will not be supposed to be timelike {\em a priori}.
The 1-form $\fstat$ has the following form, common
to all three cases:
$S^2(x) \fstat=\st_0(z) \d t +\st_2(t,x,z) \d \pp+ \st_3(t,x,z) \d z$
with $\st_0\neq 0$,
so that the condition $\fstat \wedge \d \fstat=0$ gives the following
three equations
\be
\st_{2,x}=\st_{3,x}=0, \hspace{1cm}
\st_0 \st_{2,z}+\st_2\left( \st_{3,t}-\st_{0,z}\right)-\st_3 \st_{2,t}=0.
\label{eq:sint}
\ee
Let us impose these conditions on each of the cases under study:

\subsubsubsection{Case I}

Equations (\ref{eq:sint}) imply first $L_{,x} \omega=0$.
If we take $\omega\neq 0\Rightarrow L=L_0\;(\mbox{const.})$, but since
the axis of symmetry $W_{2}$ is given by those points for which
$L(x)=0$, it follows that $L_{0}=0$ which is inconsistent with the
dimension of the spacetime, therefore it must be  $\omega=0$.

As $\omega=0\Rightarrow N=N_0\;(\mbox{const.})$, but in that case it
is easy to see that the coordinate change $\pp +bN_{0}z \mapsto \pp$
while preserves the form of the axial Killing, renders the metric in
diagonal form,
\[
\d s^2=\frac{1}{S^2(x)}\left[-\d t^2 +b^2 M^2(t) \d z^2+ \d x^2 +
L^2(x)\d \pp^2\right],
\]
which can be further transformed by suitably redefining the
coordinate $x$ to the form:
\[
\d s^2=\frac{1}{S^2(x)}\left[-\d t^2 + b^2 M^2(t) \d z^2 \right]+ \d x^2 +
L^2(x)\d \pp^2,
\]
which is that of  a (class B) warped spacetime (see \cite{CadaCo})
and can be easily seen to admit a larger group of isometries: at
least $G_{4}$ on $T_{3}$.
In this case, equations  (\ref{eq:sint}) readily imply $A=0$, and a
calculation of the modulus of $\stat$, gives
\[
(\stat\cdot \stat)=
\frac{1}{S^2}\left\{-e^{2bz}+\left(M_{,t}e^{bz}-MbB\right)^2\right\};
\]
thus, for spacetimes whose range for the $z$ coordinate
is not bounded,
we have $ (\stat\cdot \stat)>0$
when $b z\rightarrow -\infty$ unless we set $B=0$, and therefore we have
$\stat=\vec \zeta$ which is timelike in the whole manifold iff $\alpha<0$.

Therefore, the existence of a (timelike) integrable Killing vector,
implies, for this class of spacetimes, the existence of (at least) a $G_{4}$
on $T_{3}$
group of isometries which contains the original $G_{3}$ on $T_{3}$,
as well as a subgroup $G_{3}$ acting on timelike two-dimensional orbits of
constant curvature.

It is easy to see that that the Segre type of the Ricci (or Einstein)
tensor is $\{(1,1)11\}$ or some degeneracy thereof, whereas the Petrov
type of the Weyl tensor is, in general, $D$.

\subsubsubsection{Case II}

A shift in the coordinate $z$ allows us to put $B=0$ without
loss of generality.
Equations (\ref{eq:sint}) imply $A=W=F'=0$, so that $\stat=\vec \zeta$
and the metric can then be written as:
\[
\d s^2=\frac{1}{S^2(x)}\left[-\d t^2 +\exp{(-2ct)} \d z^2 \right]+ \d x^2 +
Q^2(x)\d \pp^2,
\]
where the $x$ coordinate has been re-defined in an obvious way. It
then follows that this is again a type B warped spacetime which admits a
group $G_{4}$ on $T_{3}$  of isometries which contains the $G_{3}$ on
$T_{3}$, and also as in the previous case, a subgroup $G_{3}$ acting
on timelike two-dimensional orbits of constant curvature.

As in case I, the Ricci tensor is  of the Segre type  $\{(1,1)11\}$
or some degeneracy thereof, and the Weyl tensor is type $D$.

\subsubsubsection{Case III}

Analogously to the previous case, we can put $A=0$ without
loss of generality.
In this case, however, equations (\ref{eq:sint}) imply $d=0$,
which leads to the Abelian case, thus no timelike integrable Killing
vector exists in this group. As a matter of fact, what we have just
proven is slightly more general than this; we summarize the results in
the following

\begin{propos}
Given a $G_3$ on $T_3$ group of Bianchi type II
having an Abelian $G_2$ subgroup acting orthogonally transitively and
containing an axial Killing vector, then the only integrable
Killing vectors in this group belong to the subgroup $G_{2}$.\finn
\end{propos}

This result can also be obtained for a conformal group $C_3$ containing
a $G_2$.

\vspace{5mm}
The definition of stationary (or static) cylindrically symmetric
spacetimes has been based on that for cylindrically
symmetric spacetimes.
If, on the other hand, we had started with the usual definition
of stationary axisymmetric spacetimes \cite{KRAM},
we could have imposed that the timelike Killing vector $\stat$
and the axial Killing vector $\vec \xi$ generate a $G_2$ on $T_2$ group
acting orthogonally transitively
(for non-convective
rotating fluids \cite{papapetrou,kundttrump}, see
for instance \cite{senorot}), and
the allowed Lie algebra structures would then be those four previously
discussed.
Nevertheless, it can be shown that the imposition of
orthogonal transitivity on the orbits generated
by $\stat$ and $\vec \xi$ gives no further restriction
in the {\em static} non-Abelian cases.
The conditions
\be
\bm{\xi}\wedge \bm{s} \wedge \d \bm{\xi}=0,\hspace{1cm}
\bm{\xi}\wedge \bm{s} \wedge \d \bm{s}=0
\label{eq:lastassum}
\ee
applied to each of the algebraic cases give
\bean
&\mbox{Abelian case}:&  2(F'Q-FQ')QW-(F^2W^2-Q^2)W'F=B(F'Q^2-W'F^3W)=0\\
&\mbox{Case I}:&  \mbox{Automatically satisfied}\\
&\mbox{Case II}:&  W'Q^2+F^2W^2W'-2WQQ'=F'Q^2-W'F^3W=0\\
&\mbox{Case III}:& W'=BF'=0.
\eean
Clearly,
the conditions for the existence of a timelike  integrable
Killing vector in Case II (which turns out to be $\vec\zeta$)
imply that the orbits generated by
$\stat$ and $\vec \xi$ admit 2-dimensional orthogonal surfaces
as well as the existence of (at least) a fourth linearly independent Killing
vector which, along with the previous three, generates a group
$G_{4}$ on orbits $T_{3}$.

Therefore, the assumption (\ref{eq:lastassum}) gives
no further restrictions in the non-Abelian cases
neither when imposing a timelike
integrable Killing in a geometrical sense
(i.e. including Case II),
nor in the stationary case
(when Cases II and III could be avoided).

All the above can be summarized in the following

\begin{teorem}
Given a $G_{3}$ on $T_{3}$ group that contains an orthogonally
transitive abelian $G_{2}$ subgroup generated by an axial
$\vec \xi$ and $\vec \eta$,
then it follows:
\begin{enumerate}
\item If $G_{3}$ is the maximal isometry group, then it must be
abelian.
\item If $G_{3}$ is non-abelian, then it is (locally) contained in a
$G_{4}$ on $T_{3}$, and $\vec \xi$ and $\vec \zeta$ generate an
orthogonally transitive subgroup $G_{2}$ on $T_{2}$.
\end{enumerate} \fin
\end{teorem}
Notice that, in addition, in these non-abelian cases there exist
two-dimensional timelike surfaces of constant curvature, the Segre type
of the Ricci tensor is $\{(1,1)11\}$ or some degeneracy thereof,
and the Petrov type is, in general, D.

\section*{Acknowledgements}
The authors wish to thank Marc Mars for helpful discussions.
One of us (J.C.) acknowledges financial support from NATO
(Cooperative Research Grant MA05 RF042).

\end{document}